\documentclass[aps,prl,reprint,superscriptaddress,showpacs,preprintnumbers,amsmath,amssymb]{revtex4-2}
\usepackage[dvipdfmx]{graphicx}
\usepackage{dcolumn}
\usepackage{bm}
\usepackage{amssymb, amsmath, amsfonts}
\usepackage{lipsum}
\usepackage{color}
\usepackage{multirow}
\DeclareGraphicsExtensions{{.pdf}}

\begin{document}
\newcommand{\noter}[1]{{\color{red}{#1}}}
\newcommand{\noteb}[1]{{\color{blue}{#1}}}
\newcommand{\phic}{{V_{ij}(r_{ij}^{\rm c})}}
\newcommand{\dphic}{{V^{\prime}_{ij}(r_{ij}^{\rm c})}}
\newcommand{\ddphic}{{V^{\prime\prime}_{ij}(r_{ij}^{\rm c})}}

\widetext

\title{Zero-temperature Avalanche Criticality Governing\\ Dynamical Heterogeneity in Supercooled Liquids
}

\author{Norihiro Oyama}
\affiliation{Toyota Central R\&D Labs., Inc., {Nagakute 480-1192}, Japan}

\author{Yusuke Hara}
\email{Yusuke.Hara.ys@mosk.tytlabs.co.jp}
\affiliation{Toyota Central R\&D Labs., Inc., {Nagakute 480-1192}, Japan}

\author{Takeshi Kawasaki}
\affiliation{D3 Center, The University of Osaka, Toyonaka, Osaka 560-0043, Japan}
\affiliation{Department of Physics, The University of Osaka, Toyonaka, Osaka 560-0043, Japan}
\affiliation{Department of Physics, Nagoya University, Nagoya 464-8602, Japan}

\author{Kang Kim}
\affiliation{Division of Chemical Engineering, Graduate School of Engineering Science, The University of Osaka, Toyonaka, Osaka 560-8531, Japan}

\date{\today}
\begin{abstract}
In supercooled liquids, mesoscale mobile and immobile domains are ubiquitously observed, a phenomenon known as dynamical heterogeneity. 
Extensive studies have established that the characteristic size of these domains grows upon cooling and exhibits system-size dependence. 
However, the physical origin of this domain growth remains a matter of active debate. 
In this work, using molecular simulations, we demonstrate that the temperature and system-size dependence of dynamical heterogeneity can be explained within a zero-temperature avalanche criticality picture.
\end{abstract}
\maketitle

\begin{figure}[!b]
\begin{center}
\includegraphics[width=\linewidth,angle=0]{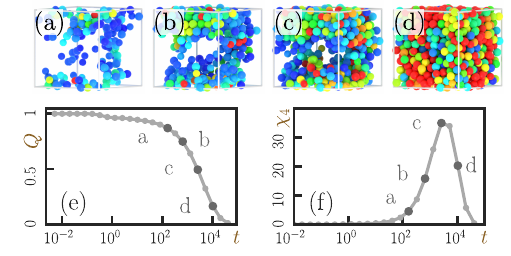}
\end{center}
\caption{
(a–d) Snapshots illustrating the growth of dynamical domains.
Each panel shows the displacement magnitude field relative to a reference configuration at the times indicated in panels (e) and (f).
Only particles with displacement magnitudes exceeding $a = 0.3$ are shown.
Warmer (cooler) colors represent larger (smaller) displacements.
(e) Semi-log plot showing the time evolution of the averaged overlap function $Q(t)$.
(f) Semi-log plot showing the time evolution of the dynamical susceptibility $\chi_4(t)$. All panels show results for $N = 1000$ and $T = 0.44$.
}
\label{fig:vis}
\end{figure}

\emph{Introduction---.}
Upon rapid cooling, various soft-matter systems~\cite{whitakerKineticStabilityHeat2015,weeksThreeDimensionalDirectImaging2000,brambillaProbingEquilibriumDynamics2009,mattssonSoftColloidsMake2009,keddieInterfaceSurfaceEffects1994,chatInfluenceTacticityGlassTransition2021,pekerHighlyProcessableMetallic1993,shengAtomicPackingShorttomediumrange2006,parryBacterialCytoplasmHas2014,nishizawaUniversalGlassformingBehavior2017,oyamaGlassyDynamicsModel2019} enter a supercooled liquid state\cite{cavagnaSupercooledLiquidsPedestrians2009}.
In this state, coexisting mobile and immobile domains are widely observed (Fig.~\ref{fig:vis}(a-d)), a phenomenon referred to as dynamical heterogeneity (DH)~\cite{trachtLengthScaleDynamic1998,freyLiquidlikeStressdrivenDynamics2025,kegelDirectObservationDynamical2000,weeksThreeDimensionalDirectImaging2000,berthierDynamicalHeterogeneitiesGlasses2011,angeliniGlasslikeDynamicsCollective2011,kanayamaRelationDynamicHeterogeneities2022}.
This behavior is quantified by a dynamical susceptibility (DS) associated with the volume of cooperatively rearranging regions~\cite{toninelliDynamicalSusceptibilityGlass2005,berthierDirectExperimentalEvidence2005}, which increases upon cooling and also exhibits finite-size effects~\cite{karmakarGrowingLengthTime2009a}.
However, the physical origin of such temperature and system-size dependence remains a matter of active debate.

A variety of theoretical approaches have been proposed to account for DH, including mode-coupling theory (MCT)~\cite{biroliDivergingLengthScale2004,biroliInhomogeneousModeCouplingTheory2006,janssenModeCouplingTheoryGlass2018}, the random first-order transition theory~\cite{bouchaudAdamGibbsKirkpatrickThirumalaiWolynesScenarioViscosity2004,kirkpatrickScalingConceptsDynamics1989}, frustration-limited domain theory~\cite{kivelsonViewpointModelTheory2013}, the distinguishable-particle lattice model~\cite{zhangEmergentFacilitationBehavior2017,lulliKovacsEffectGlass2021,leeFragileGlassesAssociated2020,lulliSpatialHeterogeneitiesStructural2020,leeLargeHeatcapacityJump2021}, and approaches based on structural order parameters~\cite{tongRevealingHiddenStructural2018,tongStructuralOrderGenuine2019}. 
In this work, we focus in particular on dynamical facilitation.
This concept originates from kinetically constrained models~\cite{fredricksonKineticIsingModel1984,jckleHierarchicallyConstrainedKinetic1991,kobKineticLatticegasModel1993} and seeks to explain nontrivial glassy dynamics through avalanche-like dynamical rules, in which the occurrence of a local structural relaxation event facilitates subsequent relaxations in its surroundings~\cite{chackoElastoplasticityMediatesDynamical2021a,scallietThirtyMillisecondsLife2022}.
Recently, Ref.~\cite{tahaeiScalingDescriptionDynamical2023a} demonstrated that, in a lattice model called the thermal elastoplastic model (t-EPM), the DS exhibits temperature and system-size dependences similar to those observed in molecular simulations~\cite{karmakarGrowingLengthTime2009a}. Moreover, such parameter dependences are well described by a zero-temperature (thermal) avalanche criticality picture analogous to plastic deformations in sheared athermal glasses~\cite{oyamaUnifiedViewAvalanche2021,oyamaInstantaneousNormalModes2021b,oyamaScaleSeparationShearInduced2024a,oyamaShearinducedCriticalityGlasses2023a,linScalingDescriptionYielding2014a,ferreroElasticInterfacesDisordered2019,liuDrivingRateDependence2016}, whereas the origin of the parameter dependence of the DS in particle-based systems remains elusive.

In this Letter, we perform molecular dynamics simulations of the Kob--Andersen model (KAM)~\cite{kobTestingModecouplingTheory1995e}, a canonical model of supercooled liquids, at temperatures above $T_{\rm MCT}$, and show that the temperature and system-size dependences of the DS can be explained within the zero-temperature avalanche criticality picture.
In particular, we demonstrate the validity of the avalanche criticality picture through finite-size scaling of the DS using independently determined critical exponents.
This criticality becomes relevant only below a threshold temperature $T_{\rm ava}$, which we identify as the onset of stability enhancement.
Moreover, the breakdown of the Stokes-Einstein relation is likewise captured by avalanche criticality, consistent with Ref.~\cite{tahaeiScalingDescriptionDynamical2023a}.
These results indicate that the zero-temperature avalanche criticality picture provides a consistent description of DS in the KAM.

\emph{Simulation---.}
In this study, we perform three-dimensional ($d=3$) molecular dynamics simulations of the KAM~\cite{kobTestingModecouplingTheory1995e}, a representative model for supercooled liquids.
The KAM is a binary Lennard-Jones mixture with nonadditive interaction parameters,
which suppress crystallization down to low temperatures.
The interaction cutoff distance is set to $r_{ij}^{\rm cut}=2.5\sigma_{ij}$, where $\sigma_{ij}$ sets the interaction length scale between particles $i$ and $j$.
As our analysis involves saddle-point configurations on the potential energy landscape, the potential is smoothed at the cutoff using cubic polynomials up to the second derivative~\cite{grigeraGeometricApproachDynamic2002b,coslovichLocalizationTransitionUnderlies2019b}. 
The temperature is controlled using the Nos\'e--Hoover thermostat,
and simulations are performed in the NVT ensemble.
Full details of the simulation setup are provided in the Appendix.
The temperature $T$ is varied over the range $0.44 \le T \le 1.0$, which lies above the MCT point $T_{\rm MCT} \approx 0.435$~\cite{kobTestingModecouplingTheory1995e}.
We consider system sizes in the range $200 \le N \le 1500$.

\begin{figure}[!tb]
\begin{center}
\includegraphics[width=\linewidth,angle=0]{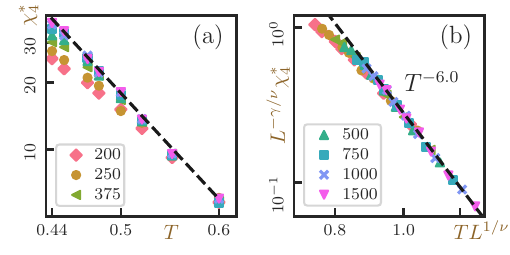}
\end{center}
\caption{
Log-log plots of the peak value of the dynamical susceptibility, $\chi_4^\ast$, as a function of $T$ in the scaling regime $T \le T_{\rm ava} \approx 0.6$.
(a) Raw data.
(b) Finite-size scaling collapse.
Symbols denote different system sizes, as indicated in the legend.
The dashed line shows the power-law scaling $\chi_4^\ast \sim T^{-\gamma}$.
}
\label{fig:chi_4}
\end{figure}

\emph{Dynamical susceptibility---.}
We first measure the DS defined as
$\chi_4(t) = N\left[\langle q^2(t)\rangle - \langle q(t)\rangle^2\right]$.
Here, $q(t)$ denotes the overlap function of A-type particles and is defined as
$q(t) = \frac{1}{N_A}\sum_{i}^{N_A}
\Theta\!\left(a - \left|\boldsymbol{r}_i(t_0+t)-\boldsymbol{r}_i(t_0)\right|\right)$.
$\Theta(x)$ denotes the Heaviside step function,
$\boldsymbol{r}_i(t)$ is the position of particle $i$ at time $t$,
and $N_A$ is the number of A-type particles.
We set $a=0.3$, which approximately corresponds to the plateau value of the mean squared displacement (MSD)~\cite{karmakarGrowingLengthTime2009a,dasCrossoverDynamicsKobAndersen2022a}.
The brackets $\langle\cdots\rangle$ denotes an average over samples and reference time.
As shown in Fig.~\ref{fig:vis}(e,f) $\chi_4(t)$ exhibits peaks at time scales where the average overlap $Q(t)\equiv\langle q(t)\rangle$ decays significantly, consistent with previous studies.
We employ the height of these peaks, $\chi_4^\ast$, as the quantitative measure of the DH and study its dependence on $T$ and $N$.

As shown in Fig.~\ref{fig:chi_4}(a), $\chi_4^\ast$ exhibits finite-size effects and a power-law dependence on $T$, implying the existence of zero-temperature criticality.
Accordingly, following the thermal avalanche criticality picture~\cite{tahaeiScalingDescriptionDynamical2023a}, we assume scaling ansatzes, $\xi \sim T^{-\nu}$ and $\chi_4^\ast \sim T^{-\gamma}$.
Within this picture, the typical linear dimension of avalanches of dynamically facilitated local rearrangements is regarded as the critical correlation length $\xi$.
The exponents $\nu$ and $\gamma$ characterize the divergence of $\xi$ and $\chi_4^\ast$, respectively.
We also introduce the fractal dimension $d_f$ through the relation
$\chi_4^\ast \sim \xi^{d_f}$ (thus, $\gamma = \nu d_f$).
If these scaling ansatzes are valid, the data for
$\chi_4^\ast L^{-\gamma/\nu}$ obtained at different system sizes
collapse onto a master curve when plotted as a function of
$T L^{1/\nu}$ (see Appendix).
Below, we determine these critical exponents $\gamma$ and $\nu$, independently.

\begin{figure}[!t]
\begin{center}
\includegraphics[width=\linewidth,angle=0]{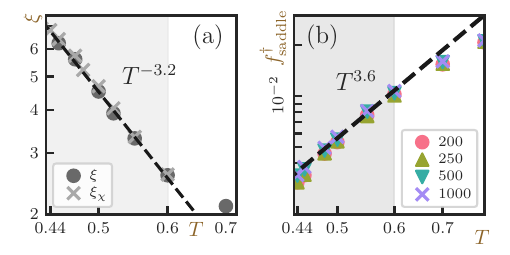}
\end{center}
\caption{
(a) Log-log plots of the two correlation lengths,
$\xi$ and $\xi_\chi$ (the latter extracted from $\chi_4^\ast$ at $N=1000$), as functions of temperature $T$.
The dashed line indicates the power-law fit to $\xi$
in the scaling regime.
(b) Log-log plots of the temperature dependence of the fraction of unstable
modes at saddle-point configurations, $f_{\rm saddle}^\dagger$.
Symbols denote different system sizes, as indicated in the legend.
The dashed line indicates the power-law fit in the scaling regime for $N = 1000$.
In both panels, the shaded region marks the scaling regime
$T \le T_{\rm ava}$.
}
\label{fig:xi}
\end{figure}

\emph{Critical correlation length---.}
As the critical correlation length $\xi$, we adopt the values of dynamical correlation length reported in 
Ref.~\cite{karmakarGrowingLengthTime2009a}.
In Ref.~\cite{karmakarGrowingLengthTime2009a}, the correlation length was
determined by applying a finite-size scaling analysis to the Binder
parameter $B \equiv
\frac{\langle [q(\tau_4)-\langle q(\tau_4)\rangle]^4\rangle}
{3\langle [q(\tau_4)-\langle q(\tau_4)\rangle]^2\rangle^2}
- 1$: 
specifically, for various combinations of $N$ and $T$, the data for $B(N,T)$ collapse onto a single master curve when plotted against the scaled system size $N/\xi^d$, with an appropriately chosen $\xi(T)$.
We emphasize that the values of $\xi$ from this analysis exhibit the same temperature dependence as the well-established dynamical correlation length estimated by the four-point structure factor $S_4(q)$~\cite{karmakarAnalysisDynamicHeterogeneity2010,tahGlassTransitionSupercooled2018a}.
In the Supplemental Material (SM), we confirmed that our simulation data for $B(N,T)$ collapse well when plotted as a function of $N/\xi^d$, using the values reported in Ref.~\cite{karmakarGrowingLengthTime2009a}.
In Fig.~\ref{fig:xi}(a), $\xi$ is plotted as a function of $T$.
In the shaded temperature regime $T \le T_{\rm ava}$, $\xi$ exhibits an apparent power-law dependence on $T$, consistent with the scaling ansatz $\xi \sim T^{-\nu}$.
A fit in this regime yields $\nu = 3.2$.
We stress that the threshold temperature distinguishing the scaling regime,
$T_{\rm ava} \approx 0.6$, is determined independently of $\xi$ according to
the criterion explained later.

\emph{Saddle mode analysis and avalanche criticality---.}
In Refs.~\cite{oyamaInstantaneousNormalModes2021b,oyamaScaleSeparationShearInduced2024a}, we demonstrated that, in zero-temperature glasses under finite-rate shear, a critical exponent associated with avalanche criticality can be extracted from the shear-rate dependence of the number of unstable instantaneous normal modes (i.e., modes with negative eigenvalues)~\cite{bembenekInstantaneousNormalModes1995,strattInstantaneousNormalModes1995}.
These normal modes are obtained as eigenmodes of the dynamical matrix, i.e., the Hessian of the potential energy with respect to particle coordinates.
In the present study, we determine the critical exponent $\gamma$ using a similar approach.
Because instantaneous normal modes at finite temperature tend to overestimate the instability of the potential energy landscape~\cite{broderixEnergyLandscapeLennardJones2000b}, we instead focus on normal modes evaluated at saddle-point configurations~\cite{angelaniSaddlesEnergyLandscape2000a,broderixEnergyLandscapeLennardJones2000b,coslovichLocalizationTransitionUnderlies2019b}.
Saddle-point configurations are obtained by minimizing the function $W \equiv (\boldsymbol{\nabla}_N E)^2$~\cite{bitzekStructuralRelaxationMade2006a,guenoleAssessmentOptimizationFast2020a}, where $\boldsymbol{\nabla}_N$ denotes the gradient with respect to all particle coordinates and $E$ is the total potential energy.

In our analysis
\cite{oyamaInstantaneousNormalModes2021b,oyamaScaleSeparationShearInduced2024a}, we consider that the number of unstable modes at saddle-point configurations, $N^\dagger_{\rm saddle}$, corresponds to the total number of local rearrangements (namely shear transformations (STs)~\cite{lerbingerRelevanceShearTransformations2022}) present in the system.
$N^\dagger_{\rm saddle}$ can be estimated as $N^\dagger_{\rm saddle} \sim N_{\rm ava} \times N_{\rm ST/ava}$.
Here, $N_{\rm ava}$ denotes the number of avalanches in the system and, from the definition of $\xi$, can be written as $N_{\rm ava} \sim (L/\xi)^d$~\cite{linScalingDescriptionYielding2014a}.
The quantity $N_{\rm ST/ava}$ represents the number of STs per avalanche and scales as $N_{\rm ST/ava} \sim \xi^{d_f}$.
Combining these relations yields $N_{\rm saddle}^\dagger\sim L^d\xi^{d_f-d}$ and we obtain for the fraction of unstable modes among all modes, $f^\dagger_{\rm saddle} \equiv N^\dagger_{\rm saddle}/(dN)$, the scaling relation $f^\dagger_{\rm saddle} \sim T^{\nu(d-d_f)}$.
Therefore, $f^\dagger_{\rm saddle}$ is expected to exhibit no system-size dependence and to obey a power law in the critical regime $T\le T_{\rm ava}$.
In Fig.~\ref{fig:xi}(b), we plot $f^\dagger_{\rm saddle}$ as a function of $T$ for several system sizes.
Consistent with the prediction above, the system-size dependence is absent, and a power-law behavior $f^\dagger_{\rm saddle} \sim T^{3.6}$ is observed for $T \le T_{\rm ava}$.
From this exponent and $\nu\approx 3.2$, we obtain $d_f = d - 3.6/\nu \approx 1.9$ and
$\gamma = \nu d_f \approx 6.0$.
Figure~\ref{fig:chi_4}(b) presents the finite-size scaling collapse obtained  using the
exponents $\nu \approx 3.2$ and $\gamma \approx 6.0$.
The observed scaling collapse supports the interpretation that, in the regime $T \le T_{\rm ava}$, the DS is governed by avalanche criticality.

We also measure another length scale $\xi_\chi$, defined as $\xi_\chi = \chi_4^{\ast,1/d_f}$.
In Fig.~\ref{fig:xi}(a), we compare $\xi$ with $\xi_\chi$, and find that $\xi_\chi$ follows the same power-law behavior as $\xi$ (their quantitative agreement is accidental due to the arbitrariness of the proportionality constant in the definition of $\xi_\chi$).
This consistency corroborates our scaling ansatzes.
We note that, as shown in the SM, if the finite-size effects of $\chi_4^\ast$ are assumed to originate from the MCT crossover, the scaling collapse of $\chi_4^\ast$ is not successful, and moreover, $\xi$ and $\xi_\chi$ are not mutually consistent.

Let us now discuss the relation between the thermal avalanche picture
proposed in Ref.~\cite{tahaeiScalingDescriptionDynamical2023a}
and our analysis based on $f_{\rm saddle}^\dagger$.
In the thermal avalanche picture, when a local relaxation releases energy, secondary events are not immediately triggered, and thus the STs forming an avalanche do not necessarily coexist simultaneously.
Instead, the facilitation through elastic interactions reduces the waiting time for subsequent relaxation events.
By contrast, in our analysis, we focus on the total number of unstable modes contained in a saddle-point configuration.
This may appear to count STs that are active simultaneously and thus to be inconsistent with the thermal avalanche picture described above.
However, this superficial discrepancy likely stems from a simplification employed in t-EPM.
Although dynamics in t-EPM is purely energetically described~\cite{tahaeiScalingDescriptionDynamical2023a}, in particulate systems, the presence of unstable modes does not necessarily mean that the structure immediately relaxes along those directions~\cite{hanggiReactionrateTheoryFifty1990,baity-jesiRevisitingConceptActivation2021}.
This can be understood as a consequence of entropic effects, as evidenced by the fact that the number of unstable modes, $N^\dagger_{\rm saddle} = dN f^\dagger_{\rm saddle}$, does not vanish even at very low temperatures, where the structural relaxation time is several orders of magnitude larger than microscopic time scales~\cite{coslovichDynamicThermodynamicCrossover2018a,dasCrossoverDynamicsKobAndersen2022a} (see Fig.~\ref{fig:xi}(b)).
Indeed, Fig.~\ref{fig:chi_4}(b) can be viewed as a demonstration of the essential consistency between our analysis and the thermal avalanche picture.

\emph{Identification of scaling regime---.}
We have determined the critical exponents from the power-law behaviors of $\xi$ and $f^\dagger_{\rm saddle}$ considering only temperatures below $T_{\rm ava} \approx 0.6$.
Here, we explain how this threshold $T_{\rm ava}$ is identified.

\begin{figure}[!tb]
\begin{center}
\includegraphics[width=\linewidth,angle=0]{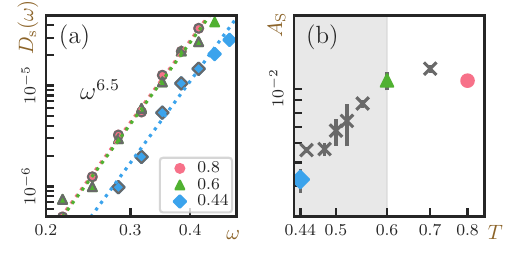}
\end{center}
\caption{
(a) Log--log plot of the low-frequency part of the vibrational density of states of stable samples, $D^{\rm S}_0(\omega)$, as a function of the eigenfrequency $\omega$.
Symbols distinguish different temperatures, as indicated in the legend.
All results are for $N=1000$.
The dotted lines represent fits to $A_{\rm S}\omega^{6.5}$.
(b) Log--log plot of $A_{\rm S}$ as a function of temperature.
The same symbols are used for the three temperatures shown in panel (a).
}
\label{fig:DoS}
\end{figure}

In the t-EPM, stability increases upon cooling, as evidenced by the temperature dependence of the probability distribution $P(x)$ of the stability parameter $x$, defined as the difference between the local stress and the local yielding threshold at each site~\cite{tahaeiScalingDescriptionDynamical2023a}.
As a corresponding quantitative measure of local stability in particulate systems, in the present study, we employ the low-frequency regime of the vibrational density of states associated with inherent structures, $D_0(\omega)$~\cite{lernerStatisticsPropertiesLowFrequency2016,mizunoContinuumLimitVibrational2017a,shimadaAnomalousVibrationalProperties2018a,kapteijnsUniversalNonphononicDensity2018,bonfantiUniversalLowFrequencyVibrational2020,richardUniversalityNonphononicVibrational2020,lernerFinitesizeEffectsNonphononic2020,xuLowfrequencyVibrationalDensity2024b}.
$D_0(\omega)$ is known to consist of quasilocalized vibrational modes (QLVs) that can trigger local plastic events~\cite{maloneyUniversalBreakdownElasticity2004, gartnerNonlinearPlasticModes2016}, and thus serve as an indicator of stability.
For small system sizes such as those considered in this study ($N \le 1500$), a temperature-independent universal power law, $D_0^{\rm S}(\omega)=A_{\rm S}\omega^{6.5}$, is observed for ensembles of mechanically stable configurations, for which the extended Hessian including boundary-condition degrees of freedom has no negative eigenvalues~\cite{xuLowfrequencyVibrationalDensity2024b}.
Figure~\ref{fig:DoS}(a) shows the measured $D_0^{\rm S}(\omega)$ for our simulations.
For all parent temperatures,
$D_0^{\rm S}(\omega)$ is consistent with the functional form $A_{\rm S}\omega^{6.5}$.
Therefore, the temperature dependence of stability
can be discussed solely in terms of the prefactor
$A_{\rm S}$.
Figure~\ref{fig:DoS}(b) shows $A_{\rm S}$
as a function of temperature.
Above $T_{\rm ava}\approx 0.6$,
$A_{\rm S}$ shows no temperature dependence,
while it decreases at lower temperatures,
indicating the enhancement of stability.
Such an increase in stability below $T_{\rm ava}$
is qualitatively consistent with the t-EPM where thermal avalanches obey zero-temperature criticality~\cite{tahaeiScalingDescriptionDynamical2023a}.
This consistency leads us to regard the range $T \le T_{\rm ava}$ as a critical regime.
The success of finite-size scaling based on the assumption of criticality in this regime (Fig.~\ref{fig:chi_4}(b)) further substantiates this interpretation.
We also emphasize that, as discussed in detail in the accompanying full paper~\cite{DHFull}, the average relaxational dynamics, $Q(t)$, also exhibits a qualitative change around the same threshold, $T_{\rm ava}\approx 0.6$.

\emph{Breakdown of SE relation---.}
In Ref.~\cite{tahaeiScalingDescriptionDynamical2023a}, two definitions of avalanche size were proposed: a lattice-based definition and an event-based one.
In the former, the avalanche size $S$ is defined as the number of lattice sites that experience local structural relaxation during an avalanche of interest, whereas in the latter, the avalanche size $\tilde{S}$ is defined as the total number of relaxation events, such that the same lattice site may be counted multiple times.
By definition, $\tilde{S} \ge S$ holds.
Critical exponents
associated with these two definitions take different values and Ref.~\cite{tahaeiScalingDescriptionDynamical2023a} proposed a scaling relation among those exponents accounting for the breakdown of the SE relation.
Since the avalanche sizes carry essentially the same information as the DS peaks, we discuss the two definitions of avalanches in the KAM in terms of the DS peaks.

\begin{figure}[!tb]
\begin{center}
\includegraphics[width=\linewidth,angle=0]{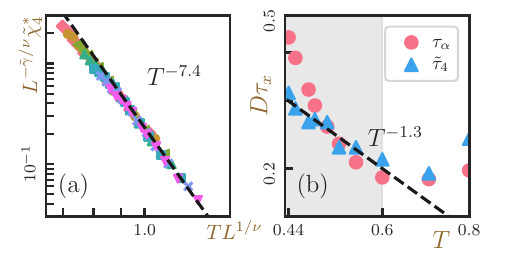}
\end{center}
\caption{
(a) Log--log plot of the displacement-based dynamical susceptibility $\tilde{\chi}_4^\ast$ as a function of temperature.
Results after finite-size scaling are shown.
Symbols have the same meaning as in Fig.~\ref{fig:chi_4}.
The dashed line represents $T^{\tilde{\gamma}}$.
(b) Log--log plot of $D\tau_x$ as a function of temperature, shown for $N=1500$.
Symbols distinguish different time scales, as indicated in the legend.
The dashed line represents the prediction of Eq.~\ref{eq:SEvio}.
}
\label{fig:tildechi4_and_SE}
\end{figure}

The susceptibility $\chi_4^\ast$ used in our analyses thus far
corresponds to the lattice-based avalanche size $S$.
As the counterpart of the event-based avalanche size $\tilde{S}$ in the KAM,
we consider a displacement-based DS,
defined as
$\tilde{\chi}_4=N\frac{\langle \Delta^2(t) \rangle-\langle \Delta(t) \rangle^2}{\langle \Delta(t) \rangle^2}$.
Here, $\Delta(t)\equiv\frac{1}{N}\sum_{i=1}^{N}\left[\boldsymbol{r}_i(t_0)-\boldsymbol{r}_i(t_0+t)\right]^2$ denotes the MSD.
With this definition, the number of local rearrangement events is reflected through variations in the magnitude of particle displacements.
Since, by definition, $\chi_4^\ast$ and $\tilde{\chi}_4^\ast$ quantify the same avalanches in different ways, they share the same correlation length $\xi$, and hence the same critical exponent $\nu$.
Defining $\tilde{d}_f$ via $\tilde{\chi}_4^\ast \sim \xi^{\tilde{d}_f}$, it follows immediately from $\tilde{\chi}_4^\ast \ge \chi_4^\ast$ that $\tilde{d}_f \ge d_f$.
Similarly, by introducing a scaling ansatz
for $\tilde{\chi}_4^\ast$ of the form
$\tilde{\chi}_4^\ast \sim T^{-\tilde{\gamma}}$
with a new critical exponent $\tilde{\gamma}$,
we obtain the scaling relation
$\tilde{\gamma} = \nu \tilde{d}_f$.
From fitting the temperature and system-size dependence
of $\tilde{\chi}_4^\ast$,
we obtain the exponent
$\tilde{\gamma} \approx 7.4$
(Fig.~\ref{fig:tildechi4_and_SE}(a);
see also the SM
for the data before scaling).
This yields
$\tilde{d}_f \approx 2.3$.
We thus confirm that
$\tilde{d}_f \ge d_f \approx 1.9$
is indeed satisfied.

We recapitulate the discussion in
Ref.~\cite{tahaeiScalingDescriptionDynamical2023a}
on how the breakdown of the SE relation
can be explained by the difference between
$d_f$ and $\tilde{d}_f$.
Over a time scale $\tau_x$ corresponding to the avalanche lifetime, the average number of local rearrangements experienced by each particle, $N_x$, scales as $N_x \sim \xi^{\tilde{d}_f}/\xi^{d_f} \sim T^{\nu(d_f-\tilde{d}_f)}$.
Assuming that, upon each local rearrangement event, the surrounding particles experience random displacements of comparable magnitude, the diffusion coefficient can be written as $D \sim N_x / \tau_x$.
Thus, the breakdown of the SE relation is expected to follow:
\begin{align}
    D\tau_x\sim N_x\sim T^{\nu(d_f-\tilde{d}_f)}.\label{eq:SEvio}
\end{align}
In Ref.~\cite{tahaeiScalingDescriptionDynamical2023a},
the $\alpha$-relaxation time $\tau_\alpha$
was adopted as the characteristic time scale $\tau_x$.
As shown in Fig.~\ref{fig:tildechi4_and_SE}(b), however, $D\tau_\alpha$ in the KAM does not follow the scaling predicted by Eq.~\eqref{eq:SEvio}.
In contrast, as demonstrated in Fig.~\ref{fig:tildechi4_and_SE}(b), when we employ the peak time of the event-based susceptibility $\tilde{\tau}_4$, defined by $\tilde{\chi}_4(\tilde{\tau}_4)=\tilde{\chi}_4^\ast$, $D\tilde{\tau}_4$ obeys the prediction of Eq.~\eqref{eq:SEvio}.
(See Appendix for procedures to determine $\tau_\alpha$ and $\tilde{\tau}_4$.)
In the derivation of Eq.~\eqref{eq:SEvio}, $\tau_x$ was implicitly assumed to represent a time scale reflecting the average number of events $N_x$ experienced by each particle.
Since $\tilde{\tau}_4$ is defined to quantify the effects of $N_x$, it is reasonable that $D\tilde{\tau}_4$ is consistent with Eq.~\eqref{eq:SEvio}.
This consistency further supports the validity of the thermal avalanche criticality picture in the KAM.

\emph{Conclusion---.}
We have shown that the zero-temperature avalanche-criticality picture can account for the parameter dependence
of the DS observed in a canonical model of supercooled liquids.
In particular, the validity of the criticality is verified by the successful finite-size scaling of the DS, using independently determined critical exponents.
We further demonstrated that, below a threshold temperature $T_{\rm ava}$, the stability of the system increases upon cooling, and that it is precisely in the temperature regime $T\le T_{\rm ava}$ that the thermal avalanche criticality becomes relevant.
Furthermore, consistent with the original article proposing the concept of thermal avalanches~\cite{tahaeiScalingDescriptionDynamical2023a}, we confirmed that the breakdown of the Stokes–Einstein relation follows a scaling relation predicted by the thermal avalanche picture (Eq.~\ref{eq:SEvio}).

We emphasize that, although our analysis in this Letter is restricted to $T > T_{\rm MCT}$, the DS is known to saturate below $T_{\rm MCT}$~\cite{coslovichDynamicThermodynamicCrossover2018a, dasCrossoverDynamicsKobAndersen2022a}.
Accordingly, the avalanche criticality picture breaks down for $T \leq T_{\rm MCT}$, and our findings do not directly imply that a glass transition, even if it exists, occurs at zero temperature.
Rather, the saturation of the DS suggests the emergence of another dominant mechanism governing glassy dynamics in deeply supercooled liquids, posing an important open problem for future work.
In our accompanying full paper~\cite{DHFull}, we propose a potential energy landscape-based interpretation of avalanche criticality that explains why the DS saturates at $T_{\rm MCT}$.
Moreover, Ref.~\cite{DHFull} presents a comprehensive comparison with prior studies, and our interpretation is consistent with this established body of knowledge.

The authors thank Yuki Takaha, Harukuni Ikeda, Hideyuki Mizuno, Atsushi Ikeda, Hajime Yoshino, and Kunimasa Miyazaki for fruitful discussions.
This work was financially supported by JSPS KAKENHI Grant Numbers JP24H02203, JP24H00192, JP25K00968.
In this research work, we used the “mdx: a platform for building data-empowered society”~\cite{suzumuraMdxCloudPlatform2022}.

\bibliography{DH_INMs}

@article{DHFull,
  title = {Potential energy landscape picture of zero-temperature avalanche criticality governing dynamics in supercooled liquids},
  author = {Oyama, Norihiro. and Hara, Yusuke and Kawasaki, Takeshi and Kim, Kang},
  journal = {},
  note = {to be submitted},
  volume = {},
  year = {},
  number = {},
  pages = {},
  doi = {},
}

@inproceedings{suzumuraMdxCloudPlatform2022,
  title = {Mdx: {{A Cloud Platform}} for {{Supporting Data Science}} and {{Cross-Disciplinary Research Collaborations}}},
  shorttitle = {Mdx},
  booktitle = {2022 {{IEEE Intl Conf}} on {{Dependable}}, {{Autonomic}} and {{Secure Computing}}, {{Intl Conf}} on {{Pervasive Intelligence}} and {{Computing}}, {{Intl Conf}} on {{Cloud}} and {{Big Data Computing}}, {{Intl Conf}} on {{Cyber Science}} and {{Technology Congress}} ({{DASC}}/{{PiCom}}/{{CBDCom}}/{{CyberSciTech}})},
  author = {Suzumura, Toyotaro and Sugiki, Akiyoshi and Takizawa, Hiroyuki and Imakura, Akira and Nakamura, Hiroshi and Taura, Kenjiro and Kudoh, Tomohiro and Hanawa, Toshihiro and Sekiya, Yuji and Kobayashi, Hiroki and Kuga, Yohei and Nakamura, Ryo and Jiang, Renhe and Kawase, Junya and Hanai, Masatoshi and Miyazaki, Hiroshi and Ishizaki, Tsutomu and Shimotoku, Daisuke and Miyamoto, Daisuke and Aida, Kento and Takefusa, Atsuko and Kurimoto, Takashi and Sasayama, Koji and Kitagawa, Naoya and Fujiwara, Ikki and Tanimura, Yusuke and Aoki, Takayuki and Endo, Toshio and Ohshima, Satoshi and Fukazawa, Keiichiro and Date, Susumu and Uchibayashi, Toshihiro},
  year = 2022,
  month = sep,
  pages = {1--7},
  publisher = {IEEE},
  address = {Falerna, Italy},
  doi = {10.1109/DASC/PiCom/CBDCom/Cy55231.2022.9927975},
  urldate = {2026-03-24},
  copyright = {https://doi.org/10.15223/policy-029},
  isbn = {978-1-6654-6297-6}
}

@article{angeliniGlasslikeDynamicsCollective2011,
  title = {Glass-like Dynamics of Collective Cell Migration},
  author = {Angelini, Thomas E. and Hannezo, Edouard and Trepat, Xavier and Marquez, Manuel and Fredberg, Jeffrey J. and Weitz, David A.},
  year = 2011,
  month = mar,
  journal = {Proceedings of the National Academy of Sciences},
  volume = {108},
  number = {12},
  pages = {4714--4719},
  issn = {0027-8424, 1091-6490},
  doi = {10.1073/pnas.1010059108},
  urldate = {2026-02-18},
  langid = {english}
}

@article{bembenekInstantaneousNormalModes1995,
  title = {Instantaneous {{Normal Modes}} and the {{Glass Transition}}},
  author = {Bembenek, Scott D. and Laird, Brian B.},
  year = 1995,
  month = feb,
  journal = {Physical Review Letters},
  volume = {74},
  number = {6},
  pages = {936--939},
  issn = {0031-9007, 1079-7114},
  doi = {10.1103/PhysRevLett.74.936},
  urldate = {2026-02-18},
  copyright = {http://link.aps.org/licenses/aps-default-license},
  langid = {english}
}

@article{biroliDivergingLengthScale2004,
  title = {Diverging Length Scale and Upper Critical Dimension in the {{Mode-Coupling Theory}} of the Glass Transition},
  author = {Biroli, G and Bouchaud, J.-P},
  year = 2004,
  month = jul,
  journal = {Europhysics Letters (EPL)},
  volume = {67},
  number = {1},
  pages = {21--27},
  issn = {0295-5075, 1286-4854},
  doi = {10.1209/epl/i2004-10044-6},
  urldate = {2026-02-18}
}

@article{kanayamaRelationDynamicHeterogeneities2022,
  title = {Relation between Dynamic Heterogeneities Observed in Scattering Experiments and Four-Body Correlations},
  author = {Kanayama, Kohsei and Hoshino, Taiki and Yamamoto, Ryoichi},
  year = 2022,
  month = apr,
  journal = {Physical Review Research},
  volume = {4},
  number = {2},
  pages = {L022006},
  issn = {2643-1564},
  doi = {10.1103/PhysRevResearch.4.L022006},
  urldate = {2026-02-18},
  langid = {english}
}

@article{scallietThirtyMillisecondsLife2022,
  title = {Thirty {{Milliseconds}} in the {{Life}} of a {{Supercooled Liquid}}},
  author = {Scalliet, Camille and Guiselin, Benjamin and Berthier, Ludovic},
  year = 2022,
  month = dec,
  journal = {Physical Review X},
  volume = {12},
  number = {4},
  pages = {041028},
  issn = {2160-3308},
  doi = {10.1103/PhysRevX.12.041028},
  urldate = {2026-02-18},
  langid = {english}
}

@article{strattInstantaneousNormalModes1995,
  title = {The {{Instantaneous Normal Modes}} of {{Liquids}}},
  author = {Stratt, Richard M.},
  year = 1995,
  month = may,
  journal = {Accounts of Chemical Research},
  volume = {28},
  number = {5},
  pages = {201--207},
  issn = {0001-4842, 1520-4898},
  doi = {10.1021/ar00053a001},
  urldate = {2026-02-18},
  langid = {english}
}

@article{angelaniSaddlesEnergyLandscape2000a,
  title = {Saddles in the {{Energy Landscape Probed}} by {{Supercooled Liquids}}},
  author = {Angelani, L. and Di Leonardo, R. and Ruocco, G. and Scala, A. and Sciortino, F.},
  year = {2000},
  month = dec,
  journal = {Physical Review Letters},
  volume = {85},
  number = {25},
  pages = {5356--5359},
  issn = {0031-9007, 1079-7114},
  doi = {10.1103/PhysRevLett.85.5356},
  urldate = {2025-10-27},
  copyright = {http://link.aps.org/licenses/aps-default-license},
  langid = {english}
}

@article{baity-jesiRevisitingConceptActivation2021,
  title = {Revisiting the Concept of Activation in Supercooled Liquids},
  author = {{Baity-Jesi}, Marco and Biroli, Giulio and Reichman, David R.},
  year = {2021},
  month = jun,
  journal = {The European Physical Journal E},
  volume = {44},
  number = {6},
  pages = {77},
  issn = {1292-8941, 1292-895X},
  doi = {10.1140/epje/s10189-021-00077-y},
  urldate = {2025-10-31},
  langid = {english}
}

@article{berthierDirectExperimentalEvidence2005,
  title = {Direct {{Experimental Evidence}} of a {{Growing Length Scale Accompanying}} the {{Glass Transition}}},
  author = {Berthier, L. and Biroli, G. and Bouchaud, J.-P. and Cipelletti, L. and Masri, D. El and L'H{\^o}te, D. and Ladieu, F. and Pierno, M.},
  year = {2005},
  month = dec,
  journal = {Science},
  volume = {310},
  number = {5755},
  pages = {1797--1800},
  issn = {0036-8075, 1095-9203},
  doi = {10.1126/science.1120714},
  urldate = {2025-10-27},
  langid = {english}
}

@book{berthierDynamicalHeterogeneitiesGlasses2011,
  title = {Dynamical {{Heterogeneities}} in {{Glasses}}, {{Colloids}}, and {{Granular Media}}},
  editor = {Berthier, Ludovic and Biroli, Giulio and Bouchaud, Jean-Philippe and Cipelletti, Luca and Van Saarloos, Wim},
  year = {2011},
  month = jul,
  publisher = {Oxford University Press},
  doi = {10.1093/acprof:oso/9780199691470.001.0001},
  urldate = {2025-10-30},
  isbn = {978-0-19-969147-0}
}

@article{biroliInhomogeneousModeCouplingTheory2006,
  title = {Inhomogeneous {{Mode-Coupling Theory}} and {{Growing Dynamic Length}} in {{Supercooled Liquids}}},
  author = {Biroli, Giulio and Bouchaud, Jean-Philippe and Miyazaki, Kunimasa and Reichman, David R.},
  year = {2006},
  month = nov,
  journal = {Physical Review Letters},
  volume = {97},
  number = {19},
  pages = {195701},
  issn = {0031-9007, 1079-7114},
  doi = {10.1103/PhysRevLett.97.195701},
  urldate = {2025-10-27},
  copyright = {http://link.aps.org/licenses/aps-default-license},
  langid = {english}
}

@article{bitzekStructuralRelaxationMade2006a,
  title = {Structural {{Relaxation Made Simple}}},
  author = {Bitzek, Erik and Koskinen, Pekka and G{\"a}hler, Franz and Moseler, Michael and Gumbsch, Peter},
  year = {2006},
  month = oct,
  journal = {Physical Review Letters},
  volume = {97},
  number = {17},
  pages = {170201},
  issn = {0031-9007, 1079-7114},
  doi = {10.1103/PhysRevLett.97.170201},
  urldate = {2025-10-31},
  copyright = {http://link.aps.org/licenses/aps-default-license},
  langid = {english}
}

@article{bonfantiUniversalLowFrequencyVibrational2020,
  title = {Universal {{Low-Frequency Vibrational Modes}} in {{Silica Glasses}}},
  author = {Bonfanti, Silvia and Guerra, Roberto and Mondal, Chandana and Procaccia, Itamar and Zapperi, Stefano},
  year = {2020},
  month = aug,
  journal = {Physical Review Letters},
  volume = {125},
  number = {8},
  pages = {085501},
  issn = {0031-9007, 1079-7114},
  doi = {10.1103/PhysRevLett.125.085501},
  urldate = {2025-10-27},
  langid = {english}
}

@article{bouchaudAdamGibbsKirkpatrickThirumalaiWolynesScenarioViscosity2004,
  title = {On the {{Adam-Gibbs-Kirkpatrick-Thirumalai-Wolynes}} Scenario for the Viscosity Increase in Glasses},
  author = {Bouchaud, Jean-Philippe and Biroli, Giulio},
  year = {2004},
  month = oct,
  journal = {The Journal of Chemical Physics},
  volume = {121},
  number = {15},
  pages = {7347--7354},
  issn = {0021-9606, 1089-7690},
  doi = {10.1063/1.1796231},
  urldate = {2025-10-27},
  langid = {english}
}

@article{brambillaProbingEquilibriumDynamics2009,
  title = {Probing the {{Equilibrium Dynamics}} of {{Colloidal Hard Spheres}} above the {{Mode-Coupling Glass Transition}}},
  author = {Brambilla, G. and El Masri, D. and Pierno, M. and Berthier, L. and Cipelletti, L. and Petekidis, G. and Schofield, A. B.},
  year = {2009},
  month = feb,
  journal = {Physical Review Letters},
  volume = {102},
  number = {8},
  pages = {085703},
  issn = {0031-9007, 1079-7114},
  doi = {10.1103/PhysRevLett.102.085703},
  urldate = {2025-10-27},
  copyright = {http://link.aps.org/licenses/aps-default-license},
  langid = {english}
}

@article{broderixEnergyLandscapeLennardJones2000b,
  title = {Energy {{Landscape}} of a {{Lennard-Jones Liquid}}: {{Statistics}} of {{Stationary Points}}},
  shorttitle = {Energy {{Landscape}} of a {{Lennard-Jones Liquid}}},
  author = {Broderix, Kurt and Bhattacharya, Kamal K. and Cavagna, Andrea and Zippelius, Annette and Giardina, Irene},
  year = {2000},
  month = dec,
  journal = {Physical Review Letters},
  volume = {85},
  number = {25},
  pages = {5360--5363},
  issn = {0031-9007, 1079-7114},
  doi = {10.1103/PhysRevLett.85.5360},
  urldate = {2025-10-27},
  copyright = {http://link.aps.org/licenses/aps-default-license},
  langid = {english}
}

@article{cavagnaSupercooledLiquidsPedestrians2009,
  title = {Supercooled Liquids for Pedestrians},
  author = {Cavagna, Andrea},
  year = {2009},
  month = jun,
  journal = {Physics Reports},
  volume = {476},
  number = {4-6},
  pages = {51--124},
  issn = {03701573},
  doi = {10.1016/j.physrep.2009.03.003},
  urldate = {2025-10-30},
  copyright = {https://www.elsevier.com/tdm/userlicense/1.0/},
  langid = {english}
}

@article{chackoElastoplasticityMediatesDynamical2021a,
  title = {Elastoplasticity {{Mediates Dynamical Heterogeneity Below}} the {{Mode Coupling Temperature}}},
  author = {Chacko, Rahul N. and Landes, Fran{\c c}ois P. and Biroli, Giulio and Dauchot, Olivier and Liu, Andrea J. and Reichman, David R.},
  year = {2021},
  month = jul,
  journal = {Physical Review Letters},
  volume = {127},
  number = {4},
  pages = {048002},
  issn = {0031-9007, 1079-7114},
  doi = {10.1103/PhysRevLett.127.048002},
  urldate = {2025-10-27},
  langid = {english}
}

@article{chatInfluenceTacticityGlassTransition2021,
  title = {Influence of {{Tacticity}} on the {{Glass-Transition Dynamics}} of {{Poly}}(Methyl Methacrylate) ({{PMMA}}) under {{Elevated Pressure}} and {{Geometrical Nanoconfinement}}},
  author = {Chat, Katarzyna and Tu, Wenkang and Beena Unni, Aparna and Adrjanowicz, Karolina},
  year = {2021},
  month = sep,
  journal = {Macromolecules},
  volume = {54},
  number = {18},
  pages = {8526--8537},
  issn = {0024-9297, 1520-5835},
  doi = {10.1021/acs.macromol.1c01341},
  urldate = {2025-10-27},
  copyright = {https://creativecommons.org/licenses/by/4.0/},
  langid = {english}
}

@article{coslovichDynamicThermodynamicCrossover2018a,
  title = {Dynamic and Thermodynamic Crossover Scenarios in the {{Kob-Andersen}} Mixture: {{Insights}} from Multi-{{CPU}} and Multi-{{GPU}} Simulations},
  shorttitle = {Dynamic and Thermodynamic Crossover Scenarios in the {{Kob-Andersen}} Mixture},
  author = {Coslovich, Daniele and Ozawa, Misaki and Kob, Walter},
  year = {2018},
  month = may,
  journal = {The European Physical Journal E},
  volume = {41},
  number = {5},
  pages = {62},
  issn = {1292-8941, 1292-895X},
  doi = {10.1140/epje/i2018-11671-2},
  urldate = {2025-10-30},
  langid = {english}
}

@article{coslovichLocalizationTransitionUnderlies2019b,
  title = {A Localization Transition Underlies the Mode-Coupling Crossover of Glasses},
  author = {Coslovich, Daniele and Ninarello, Andrea and Berthier, Ludovic},
  year = {2019},
  month = dec,
  journal = {SciPost Physics},
  volume = {7},
  number = {6},
  pages = {077},
  issn = {2542-4653},
  doi = {10.21468/SciPostPhys.7.6.077},
  urldate = {2025-10-27},
  langid = {english}
}

@article{dasCrossoverDynamicsKobAndersen2022a,
  title = {Crossover in Dynamics in the {{Kob-Andersen}} Binary Mixture Glass-Forming Liquid},
  author = {Das, Pallabi and Sastry, Srikanth},
  year = {2022},
  month = jun,
  journal = {Journal of Non-Crystalline Solids: X},
  volume = {14},
  pages = {100098},
  issn = {25901591},
  doi = {10.1016/j.nocx.2022.100098},
  urldate = {2025-10-27},
  langid = {english}
}

@article{ferreroElasticInterfacesDisordered2019,
  title = {Elastic {{Interfaces}} on {{Disordered Substrates}}: {{From Mean-Field Depinning}} to {{Yielding}}},
  shorttitle = {Elastic {{Interfaces}} on {{Disordered Substrates}}},
  author = {Ferrero, E. E. and Jagla, E. A.},
  year = {2019},
  month = nov,
  journal = {Physical Review Letters},
  volume = {123},
  number = {21},
  pages = {218002},
  issn = {0031-9007, 1079-7114},
  doi = {10.1103/PhysRevLett.123.218002},
  urldate = {2025-10-27},
  langid = {english}
}

@article{fredricksonKineticIsingModel1984,
  title = {Kinetic {{Ising Model}} of the {{Glass Transition}}},
  author = {Fredrickson, Glenn H. and Andersen, Hans C.},
  year = {1984},
  month = sep,
  journal = {Physical Review Letters},
  volume = {53},
  number = {13},
  pages = {1244--1247},
  issn = {0031-9007},
  doi = {10.1103/PhysRevLett.53.1244},
  urldate = {2025-10-27},
  copyright = {http://link.aps.org/licenses/aps-default-license},
  langid = {english}
}

@article{freyLiquidlikeStressdrivenDynamics2025,
  title = {Liquid-like versus Stress-Driven Dynamics in a Metallic Glass Former Observed by Temperature Scanning {{X-ray}} Photon Correlation Spectroscopy},
  author = {Frey, Maximilian and Neuber, Nico and Riegler, Sascha Sebastian and Cornet, Antoine and Chushkin, Yuriy and Zontone, Federico and Ruschel, Lucas Matthias and Adam, Bastian and Nabahat, Mehran and Yang, Fan and Shen, Jie and Westermeier, Fabian and Sprung, Michael and Cangialosi, Daniele and Di Lisio, Valerio and Gallino, Isabella and Busch, Ralf and Ruta, Beatrice and Pineda, Eloi},
  year = {2025},
  month = may,
  journal = {Nature Communications},
  volume = {16},
  number = {1},
  pages = {4429},
  issn = {2041-1723},
  doi = {10.1038/s41467-025-59767-2},
  urldate = {2025-10-27},
  langid = {english}
}

@article{gartnerNonlinearPlasticModes2016,
  title = {Nonlinear Plastic Modes in Disordered Solids},
  author = {Gartner, Luka and Lerner, Edan},
  year = {2016},
  month = jan,
  journal = {Physical Review E},
  volume = {93},
  number = {1},
  pages = {011001},
  issn = {2470-0045, 2470-0053},
  doi = {10.1103/PhysRevE.93.011001},
  urldate = {2025-10-27},
  copyright = {http://link.aps.org/licenses/aps-default-license},
  langid = {english}
}

@article{grigeraGeometricApproachDynamic2002b,
  title = {Geometric {{Approach}} to the {{Dynamic Glass Transition}}},
  author = {Grigera, Tom{\'a}s S. and Cavagna, Andrea and Giardina, Irene and Parisi, Giorgio},
  year = {2002},
  month = jan,
  journal = {Physical Review Letters},
  volume = {88},
  number = {5},
  pages = {055502},
  issn = {0031-9007, 1079-7114},
  doi = {10.1103/PhysRevLett.88.055502},
  urldate = {2025-10-27},
  copyright = {http://link.aps.org/licenses/aps-default-license},
  langid = {english}
}

@article{guenoleAssessmentOptimizationFast2020a,
  title = {Assessment and Optimization of the Fast Inertial Relaxation Engine (Fire) for Energy Minimization in Atomistic Simulations and Its Implementation in Lammps},
  author = {Gu{\'e}nol{\'e}, Julien and N{\"o}hring, Wolfram G. and Vaid, Aviral and Houll{\'e}, Fr{\'e}d{\'e}ric and Xie, Zhuocheng and Prakash, Aruna and Bitzek, Erik},
  year = {2020},
  month = apr,
  journal = {Computational Materials Science},
  volume = {175},
  pages = {109584},
  issn = {09270256},
  doi = {10.1016/j.commatsci.2020.109584},
  urldate = {2025-10-31},
  langid = {english}
}

@article{hanggiReactionrateTheoryFifty1990,
  title = {Reaction-Rate Theory: Fifty Years after {{Kramers}}},
  shorttitle = {Reaction-Rate Theory},
  author = {H{\"a}nggi, Peter and Talkner, Peter and Borkovec, Michal},
  year = {1990},
  month = apr,
  journal = {Reviews of Modern Physics},
  volume = {62},
  number = {2},
  pages = {251--341},
  issn = {0034-6861, 1539-0756},
  doi = {10.1103/RevModPhys.62.251},
  urldate = {2025-10-27},
  copyright = {http://link.aps.org/licenses/aps-default-license},
  langid = {english}
}

@article{ingebrigtsenCrystallizationInstabilityGlassForming2019a,
  title = {Crystallization {{Instability}} in {{Glass-Forming Mixtures}}},
  author = {Ingebrigtsen, Trond S. and Dyre, Jeppe C. and Schr{\o}der, Thomas B. and Royall, C. Patrick},
  year = {2019},
  month = aug,
  journal = {Physical Review X},
  volume = {9},
  number = {3},
  pages = {031016},
  issn = {2160-3308},
  doi = {10.1103/PhysRevX.9.031016},
  urldate = {2025-10-27},
  langid = {english}
}

@article{janssenModeCouplingTheoryGlass2018,
  title = {Mode-{{Coupling Theory}} of the {{Glass Transition}}: {{A Primer}}},
  shorttitle = {Mode-{{Coupling Theory}} of the {{Glass Transition}}},
  author = {Janssen, Liesbeth M. C.},
  year = {2018},
  month = oct,
  journal = {Frontiers in Physics},
  volume = {6},
  pages = {97},
  issn = {2296-424X},
  doi = {10.3389/fphy.2018.00097},
  urldate = {2025-10-27},
  langid = {english}
}

@article{jckleHierarchicallyConstrainedKinetic1991,
  title = {A Hierarchically Constrained Kinetic {{Ising}} Model},
  author = {J{\"a}ckle, J. and Eisinger, S.},
  year = {1991},
  month = feb,
  journal = {Zeitschrift f{\"u}r Physik B Condensed Matter},
  volume = {84},
  number = {1},
  pages = {115--124},
  issn = {0722-3277, 1434-6036},
  doi = {10.1007/BF01453764},
  urldate = {2025-10-30},
  copyright = {http://www.springer.com/tdm},
  langid = {english}
}

@article{kapteijnsUniversalNonphononicDensity2018,
  title = {Universal {{Nonphononic Density}} of {{States}} in {{2D}}, {{3D}}, and {{4D Glasses}}},
  author = {Kapteijns, Geert and Bouchbinder, Eran and Lerner, Edan},
  year = {2018},
  month = aug,
  journal = {Physical Review Letters},
  volume = {121},
  number = {5},
  pages = {055501},
  issn = {0031-9007, 1079-7114},
  doi = {10.1103/PhysRevLett.121.055501},
  urldate = {2025-10-27},
  langid = {english}
}

@article{karmakarAnalysisDynamicHeterogeneity2010,
  title = {Analysis of {{Dynamic Heterogeneity}} in a {{Glass Former}} from the {{Spatial Correlations}} of {{Mobility}}},
  author = {Karmakar, Smarajit and Dasgupta, Chandan and Sastry, Srikanth},
  year = {2010},
  month = jul,
  journal = {Physical Review Letters},
  volume = {105},
  number = {1},
  pages = {015701},
  issn = {0031-9007, 1079-7114},
  doi = {10.1103/PhysRevLett.105.015701},
  urldate = {2025-10-31},
  copyright = {http://link.aps.org/licenses/aps-default-license},
  langid = {english}
}

@article{karmakarGrowingLengthTime2009a,
  title = {Growing Length and Time Scales in Glass-Forming Liquids},
  author = {Karmakar, Smarajit and Dasgupta, Chandan and Sastry, Srikanth},
  year = {2009},
  month = mar,
  journal = {Proceedings of the National Academy of Sciences},
  volume = {106},
  number = {10},
  pages = {3675--3679},
  issn = {0027-8424, 1091-6490},
  doi = {10.1073/pnas.0811082106},
  urldate = {2025-10-27},
  langid = {english}
}

@article{keddieInterfaceSurfaceEffects1994,
  title = {Interface and Surface Effects on the Glass-Transition Temperature in Thin Polymer Films},
  author = {Keddie, Joseph L. and Jones, Richard A. L. and Cory, Rachel A.},
  year = {1994},
  journal = {Faraday Discussions},
  volume = {98},
  pages = {219},
  issn = {1359-6640, 1364-5498},
  doi = {10.1039/fd9949800219},
  urldate = {2025-10-30},
  langid = {english}
}

@article{kegelDirectObservationDynamical2000,
  title = {Direct {{Observation}} of {{Dynamical Heterogeneities}} in {{Colloidal Hard-Sphere Suspensions}}},
  author = {Kegel, Willem K. and Van Blaaderen, Alfons},
  year = {2000},
  month = jan,
  journal = {Science},
  volume = {287},
  number = {5451},
  pages = {290--293},
  issn = {0036-8075, 1095-9203},
  doi = {10.1126/science.287.5451.290},
  urldate = {2025-10-27},
  langid = {english}
}

@article{kirkpatrickScalingConceptsDynamics1989,
  title = {Scaling Concepts for the Dynamics of Viscous Liquids near an Ideal Glassy State},
  author = {Kirkpatrick, T. R. and Thirumalai, D. and Wolynes, P. G.},
  year = {1989},
  month = jul,
  journal = {Physical Review A},
  volume = {40},
  number = {2},
  pages = {1045--1054},
  issn = {0556-2791},
  doi = {10.1103/PhysRevA.40.1045},
  urldate = {2025-10-27},
  copyright = {http://link.aps.org/licenses/aps-default-license},
  langid = {english}
}

@article{kivelsonViewpointModelTheory2013,
  title = {A {{Viewpoint}}, {{Model}} and {{Theory}} for {{Supercooled Liquids}}},
  author = {Kivelson, D. and Tarjus, G. and Kivelson, S. A.},
  year = {2013},
  month = may,
  journal = {Progress of Theoretical Physics Supplement},
  volume = {126},
  number = {0},
  pages = {289--299},
  issn = {0375-9687},
  doi = {10.1143/PTP.126.289},
  urldate = {2025-10-31},
  langid = {english}
}

@article{kobKineticLatticegasModel1993,
  title = {Kinetic Lattice-Gas Model of Cage Effects in High-Density Liquids and a Test of Mode-Coupling Theory of the Ideal-Glass Transition},
  author = {Kob, Walter and Andersen, Hans C.},
  year = {1993},
  month = dec,
  journal = {Physical Review E},
  volume = {48},
  number = {6},
  pages = {4364--4377},
  issn = {1063-651X, 1095-3787},
  doi = {10.1103/PhysRevE.48.4364},
  urldate = {2025-10-27},
  copyright = {http://link.aps.org/licenses/aps-default-license},
  langid = {english}
}

@article{kobTestingModecouplingTheory1995e,
  title = {Testing Mode-Coupling Theory for a Supercooled Binary {{Lennard-Jones}} Mixture. {{II}}. {{Intermediate}} Scattering Function and Dynamic Susceptibility},
  author = {Kob, Walter and Andersen, Hans C.},
  year = {1995},
  month = oct,
  journal = {Physical Review E},
  volume = {52},
  number = {4},
  pages = {4134--4153},
  issn = {1063-651X, 1095-3787},
  doi = {10.1103/PhysRevE.52.4134},
  urldate = {2025-10-27},
  copyright = {http://link.aps.org/licenses/aps-default-license},
  langid = {english}
}

@article{leeFragileGlassesAssociated2020,
  title = {Fragile {{Glasses Associated}} with a {{Dramatic Drop}} of {{Entropy}} under {{Supercooling}}},
  author = {Lee, Chun-Shing and Lulli, Matteo and Zhang, Ling-Han and Deng, Hai-Yao and Lam, Chi-Hang},
  year = {2020},
  month = dec,
  journal = {Physical Review Letters},
  volume = {125},
  number = {26},
  pages = {265703},
  issn = {0031-9007, 1079-7114},
  doi = {10.1103/PhysRevLett.125.265703},
  urldate = {2025-10-27},
  langid = {english}
}

@article{leeLargeHeatcapacityJump2021,
  title = {Large Heat-Capacity Jump in Cooling-Heating of Fragile Glass from Kinetic {{Monte Carlo}} Simulations Based on a Two-State Picture},
  author = {Lee, Chun-Shing and Deng, Hai-Yao and Yip, Cho-Tung and Lam, Chi-Hang},
  year = {2021},
  month = aug,
  journal = {Physical Review E},
  volume = {104},
  number = {2},
  pages = {024131},
  issn = {2470-0045, 2470-0053},
  doi = {10.1103/PhysRevE.104.024131},
  urldate = {2025-10-27},
  langid = {english}
}

@article{lerbingerRelevanceShearTransformations2022,
  title = {Relevance of {{Shear Transformations}} in the {{Relaxation}} of {{Supercooled Liquids}}},
  author = {Lerbinger, Matthias and Barbot, Armand and Vandembroucq, Damien and Patinet, Sylvain},
  year = {2022},
  month = oct,
  journal = {Physical Review Letters},
  volume = {129},
  number = {19},
  pages = {195501},
  issn = {0031-9007, 1079-7114},
  doi = {10.1103/PhysRevLett.129.195501},
  urldate = {2025-10-27},
  langid = {english}
}

@article{lernerFinitesizeEffectsNonphononic2020,
  title = {Finite-Size Effects in the Nonphononic Density of States in Computer Glasses},
  author = {Lerner, Edan},
  year = {2020},
  month = mar,
  journal = {Physical Review E},
  volume = {101},
  number = {3},
  pages = {032120},
  issn = {2470-0045, 2470-0053},
  doi = {10.1103/PhysRevE.101.032120},
  urldate = {2025-10-27},
  langid = {english}
}

@article{lernerStatisticsPropertiesLowFrequency2016,
  title = {Statistics and {{Properties}} of {{Low-Frequency Vibrational Modes}} in {{Structural Glasses}}},
  author = {Lerner, Edan and D{\"u}ring, Gustavo and Bouchbinder, Eran},
  year = {2016},
  month = jul,
  journal = {Physical Review Letters},
  volume = {117},
  number = {3},
  pages = {035501},
  issn = {0031-9007, 1079-7114},
  doi = {10.1103/PhysRevLett.117.035501},
  urldate = {2025-10-27},
  copyright = {http://link.aps.org/licenses/aps-default-license},
  langid = {english}
}

@article{linScalingDescriptionYielding2014a,
  title = {Scaling Description of the Yielding Transition in Soft Amorphous Solids at Zero Temperature},
  author = {Lin, Jie and Lerner, Edan and Rosso, Alberto and Wyart, Matthieu},
  year = {2014},
  month = oct,
  journal = {Proceedings of the National Academy of Sciences},
  volume = {111},
  number = {40},
  pages = {14382--14387},
  issn = {0027-8424, 1091-6490},
  doi = {10.1073/pnas.1406391111},
  urldate = {2025-10-27},
  langid = {english}
}

@article{liuDrivingRateDependence2016,
  title = {Driving {{Rate Dependence}} of {{Avalanche Statistics}} and {{Shapes}} at the {{Yielding Transition}}},
  author = {Liu, Chen and Ferrero, Ezequiel E. and Puosi, Francesco and Barrat, Jean-Louis and Martens, Kirsten},
  year = {2016},
  month = feb,
  journal = {Physical Review Letters},
  volume = {116},
  number = {6},
  pages = {065501},
  issn = {0031-9007, 1079-7114},
  doi = {10.1103/PhysRevLett.116.065501},
  urldate = {2025-10-27},
  copyright = {http://link.aps.org/licenses/aps-default-license},
  langid = {english}
}

@article{lulliKovacsEffectGlass2021,
  title = {Kovacs Effect in Glass with Material Memory Revealed in Non-Equilibrium Particle Interactions},
  author = {Lulli, Matteo and Lee, Chun-Shing and Zhang, Ling-Han and Deng, Hai-Yao and Lam, Chi-Hang},
  year = {2021},
  month = sep,
  journal = {Journal of Statistical Mechanics: Theory and Experiment},
  volume = {2021},
  number = {9},
  pages = {093303},
  issn = {1742-5468},
  doi = {10.1088/1742-5468/ac1f26},
  urldate = {2025-10-27},
  langid = {english}
}

@article{lulliSpatialHeterogeneitiesStructural2020,
  title = {Spatial {{Heterogeneities}} in {{Structural Temperature Cause Kovacs}}' {{Expansion Gap Paradox}} in {{Aging}} of {{Glasses}}},
  author = {Lulli, Matteo and Lee, Chun-Shing and Deng, Hai-Yao and Yip, Cho-Tung and Lam, Chi-Hang},
  year = {2020},
  month = mar,
  journal = {Physical Review Letters},
  volume = {124},
  number = {9},
  pages = {095501},
  issn = {0031-9007, 1079-7114},
  doi = {10.1103/PhysRevLett.124.095501},
  urldate = {2025-10-27},
  langid = {english}
}

@article{maloneyUniversalBreakdownElasticity2004,
  title = {Universal {{Breakdown}} of {{Elasticity}} at the {{Onset}} of {{Material Failure}}},
  author = {Maloney, Craig and Lema{\^i}tre, Ana{\"e}l},
  year = {2004},
  month = nov,
  journal = {Physical Review Letters},
  volume = {93},
  number = {19},
  pages = {195501},
  issn = {0031-9007, 1079-7114},
  doi = {10.1103/PhysRevLett.93.195501},
  urldate = {2025-10-27},
  copyright = {http://link.aps.org/licenses/aps-default-license},
  langid = {english}
}

@article{mattssonSoftColloidsMake2009,
  title = {Soft Colloids Make Strong Glasses},
  author = {Mattsson, Johan and Wyss, Hans M. and {Fernandez-Nieves}, Alberto and Miyazaki, Kunimasa and Hu, Zhibing and Reichman, David R. and Weitz, David A.},
  year = {2009},
  month = nov,
  journal = {Nature},
  volume = {462},
  number = {7269},
  pages = {83--86},
  issn = {0028-0836, 1476-4687},
  doi = {10.1038/nature08457},
  urldate = {2025-10-27},
  copyright = {http://www.springer.com/tdm},
  langid = {english}
}

@article{mizunoContinuumLimitVibrational2017a,
  title = {Continuum Limit of the Vibrational Properties of Amorphous Solids},
  author = {Mizuno, Hideyuki and Shiba, Hayato and Ikeda, Atsushi},
  year = {2017},
  month = nov,
  journal = {Proceedings of the National Academy of Sciences},
  volume = {114},
  number = {46},
  issn = {0027-8424, 1091-6490},
  doi = {10.1073/pnas.1709015114},
  urldate = {2025-10-27},
  langid = {english}
}

@article{nishizawaUniversalGlassformingBehavior2017,
  title = {Universal Glass-Forming Behavior of in Vitro and Living Cytoplasm},
  author = {Nishizawa, Kenji and Fujiwara, Kei and Ikenaga, Masahiro and Nakajo, Nobushige and Yanagisawa, Miho and Mizuno, Daisuke},
  year = {2017},
  month = nov,
  journal = {Scientific Reports},
  volume = {7},
  number = {1},
  pages = {15143},
  issn = {2045-2322},
  doi = {10.1038/s41598-017-14883-y},
  urldate = {2025-10-27},
  langid = {english}
}

@article{ortliebProbingExcitationsCooperatively2023,
  title = {Probing Excitations and Cooperatively Rearranging Regions in Deeply Supercooled Liquids},
  author = {Ortlieb, Levke and Ingebrigtsen, Trond S. and Hallett, James E. and Turci, Francesco and Royall, C. Patrick},
  year = {2023},
  month = may,
  journal = {Nature Communications},
  volume = {14},
  number = {1},
  pages = {2621},
  issn = {2041-1723},
  doi = {10.1038/s41467-023-37793-2},
  urldate = {2025-10-27},
  langid = {english}
}

@article{oyamaGlassyDynamicsModel2019,
  title = {Glassy Dynamics of a Model of Bacterial Cytoplasm with Metabolic Activities},
  author = {Oyama, Norihiro and Kawasaki, Takeshi and Mizuno, Hideyuki and Ikeda, Atsushi},
  year = {2019},
  month = dec,
  journal = {Physical Review Research},
  volume = {1},
  number = {3},
  pages = {032038},
  issn = {2643-1564},
  doi = {10.1103/PhysRevResearch.1.032038},
  urldate = {2025-10-27},
  langid = {english}
}

@article{oyamaInstantaneousNormalModes2021b,
  title = {Instantaneous {{Normal Modes Reveal Structural Signatures}} for the {{Herschel-Bulkley Rheology}} in {{Sheared Glasses}}},
  author = {Oyama, Norihiro and Mizuno, Hideyuki and Ikeda, Atsushi},
  year = {2021},
  month = sep,
  journal = {Physical Review Letters},
  volume = {127},
  number = {10},
  pages = {108003},
  issn = {0031-9007, 1079-7114},
  doi = {10.1103/PhysRevLett.127.108003},
  urldate = {2025-10-27},
  langid = {english}
}

@article{oyamaScaleSeparationShearInduced2024a,
  title = {Scale {{Separation}} of {{Shear-Induced Criticality}} in {{Glasses}}},
  author = {Oyama, Norihiro and Kawasaki, Takeshi and Kim, Kang and Mizuno, Hideyuki},
  year = {2024},
  month = apr,
  journal = {Physical Review Letters},
  volume = {132},
  number = {14},
  pages = {148201},
  issn = {0031-9007, 1079-7114},
  doi = {10.1103/PhysRevLett.132.148201},
  urldate = {2025-10-27},
  langid = {english}
}

@article{oyamaShearinducedCriticalityGlasses2023a,
  title = {Shear-Induced Criticality in Glasses Shares Qualitative Similarities with the {{Gardner}} Phase},
  author = {Oyama, Norihiro and Mizuno, Hideyuki and Ikeda, Atsushi},
  year = {2023},
  journal = {Soft Matter},
  volume = {19},
  number = {32},
  pages = {6074--6087},
  issn = {1744-683X, 1744-6848},
  doi = {10.1039/D3SM00512G},
  urldate = {2025-10-27},
  langid = {english}
}

@article{oyamaUnifiedViewAvalanche2021,
  title = {Unified View of Avalanche Criticality in Sheared Glasses},
  author = {Oyama, Norihiro and Mizuno, Hideyuki and Ikeda, Atsushi},
  year = {2021},
  month = jul,
  journal = {Physical Review E},
  volume = {104},
  number = {1},
  pages = {015002},
  issn = {2470-0045, 2470-0053},
  doi = {10.1103/PhysRevE.104.015002},
  urldate = {2025-10-27},
  langid = {english}
}

@article{parryBacterialCytoplasmHas2014,
  title = {The {{Bacterial Cytoplasm Has Glass-like Properties}} and {{Is Fluidized}} by {{Metabolic Activity}}},
  author = {Parry, Bradley~R. and Surovtsev, Ivan~V. and Cabeen, Matthew~T. and O'Hern, Corey~S. and Dufresne, Eric~R. and {Jacobs-Wagner}, Christine},
  year = {2014},
  month = jan,
  journal = {Cell},
  volume = {156},
  number = {1-2},
  pages = {183--194},
  issn = {00928674},
  doi = {10.1016/j.cell.2013.11.028},
  urldate = {2025-10-27},
  langid = {english}
}

@article{pekerHighlyProcessableMetallic1993,
  title = {A Highly Processable Metallic Glass: {{Zr41}}.{{2Ti13}}.{{8Cu12}}.{{5Ni10}}.{{0Be22}}.5},
  shorttitle = {A Highly Processable Metallic Glass},
  author = {Peker, A. and Johnson, W. L.},
  year = {1993},
  month = oct,
  journal = {Applied Physics Letters},
  volume = {63},
  number = {17},
  pages = {2342--2344},
  issn = {0003-6951, 1077-3118},
  doi = {10.1063/1.110520},
  urldate = {2025-10-27},
  langid = {english}
}

@article{richardUniversalityNonphononicVibrational2020,
  title = {Universality of the {{Nonphononic Vibrational Spectrum}} across {{Different Classes}} of {{Computer Glasses}}},
  author = {Richard, David and {Gonz{\'a}lez-L{\'o}pez}, Karina and Kapteijns, Geert and Pater, Robert and Vaknin, Talya and Bouchbinder, Eran and Lerner, Edan},
  year = {2020},
  month = aug,
  journal = {Physical Review Letters},
  volume = {125},
  number = {8},
  pages = {085502},
  issn = {0031-9007, 1079-7114},
  doi = {10.1103/PhysRevLett.125.085502},
  urldate = {2025-10-27},
  langid = {english}
}

@article{sastrySignaturesDistinctDynamical1998b,
  title = {Signatures of Distinct Dynamical Regimes in the Energy Landscape of a Glass-Forming Liquid},
  author = {Sastry, Srikanth and Debenedetti, Pablo G. and Stillinger, Frank H.},
  year = {1998},
  month = jun,
  journal = {Nature},
  volume = {393},
  number = {6685},
  pages = {554--557},
  issn = {0028-0836, 1476-4687},
  doi = {10.1038/31189},
  urldate = {2025-10-27},
  copyright = {http://www.springer.com/tdm},
  langid = {english}
}

@article{shengAtomicPackingShorttomediumrange2006,
  title = {Atomic Packing and Short-to-Medium-Range Order in Metallic Glasses},
  author = {Sheng, H. W. and Luo, W. K. and Alamgir, F. M. and Bai, J. M. and Ma, E.},
  year = {2006},
  month = jan,
  journal = {Nature},
  volume = {439},
  number = {7075},
  pages = {419--425},
  issn = {0028-0836, 1476-4687},
  doi = {10.1038/nature04421},
  urldate = {2025-10-27},
  copyright = {http://www.springer.com/tdm},
  langid = {english}
}

@article{shimadaAnomalousVibrationalProperties2018a,
  title = {Anomalous Vibrational Properties in the Continuum Limit of Glasses},
  author = {Shimada, Masanari and Mizuno, Hideyuki and Ikeda, Atsushi},
  year = {2018},
  month = feb,
  journal = {Physical Review E},
  volume = {97},
  number = {2},
  pages = {022609},
  issn = {2470-0045, 2470-0053},
  doi = {10.1103/PhysRevE.97.022609},
  urldate = {2025-10-27},
  langid = {english}
}

@article{tahaeiScalingDescriptionDynamical2023a,
  title = {Scaling {{Description}} of {{Dynamical Heterogeneity}} and {{Avalanches}} of {{Relaxation}} in {{Glass-Forming Liquids}}},
  author = {Tahaei, Ali and Biroli, Giulio and Ozawa, Misaki and Popovi{\'c}, Marko and Wyart, Matthieu},
  year = {2023},
  month = sep,
  journal = {Physical Review X},
  volume = {13},
  number = {3},
  pages = {031034},
  issn = {2160-3308},
  doi = {10.1103/PhysRevX.13.031034},
  urldate = {2025-10-27},
  langid = {english}
}

@article{tahGlassTransitionSupercooled2018a,
  title = {Glass {{Transition}} in {{Supercooled Liquids}} with {{Medium-Range Crystalline Order}}},
  author = {Tah, Indrajit and Sengupta, Shiladitya and Sastry, Srikanth and Dasgupta, Chandan and Karmakar, Smarajit},
  year = {2018},
  month = aug,
  journal = {Physical Review Letters},
  volume = {121},
  number = {8},
  pages = {085703},
  issn = {0031-9007, 1079-7114},
  doi = {10.1103/PhysRevLett.121.085703},
  urldate = {2025-10-27},
  langid = {english}
}

@article{tongRevealingHiddenStructural2018,
  title = {Revealing {{Hidden Structural Order Controlling Both Fast}} and {{Slow Glassy Dynamics}} in {{Supercooled Liquids}}},
  author = {Tong, Hua and Tanaka, Hajime},
  year = {2018},
  month = mar,
  journal = {Physical Review X},
  volume = {8},
  number = {1},
  pages = {011041},
  issn = {2160-3308},
  doi = {10.1103/PhysRevX.8.011041},
  urldate = {2025-10-27},
  langid = {english}
}

@article{tongStructuralOrderGenuine2019,
  title = {Structural Order as a Genuine Control Parameter of Dynamics in Simple Glass Formers},
  author = {Tong, Hua and Tanaka, Hajime},
  year = {2019},
  month = dec,
  journal = {Nature Communications},
  volume = {10},
  number = {1},
  pages = {5596},
  issn = {2041-1723},
  doi = {10.1038/s41467-019-13606-3},
  urldate = {2025-10-27},
  langid = {english}
}

@article{toninelliDynamicalSusceptibilityGlass2005,
  title = {Dynamical Susceptibility of Glass Formers: {{Contrasting}} the Predictions of Theoretical Scenarios},
  shorttitle = {Dynamical Susceptibility of Glass Formers},
  author = {Toninelli, Cristina and Wyart, Matthieu and Berthier, Ludovic and Biroli, Giulio and Bouchaud, Jean-Philippe},
  year = {2005},
  month = apr,
  journal = {Physical Review E},
  volume = {71},
  number = {4},
  pages = {041505},
  issn = {1539-3755, 1550-2376},
  doi = {10.1103/PhysRevE.71.041505},
  urldate = {2025-10-27},
  copyright = {http://link.aps.org/licenses/aps-default-license},
  langid = {english}
}

@article{trachtLengthScaleDynamic1998,
  title = {Length {{Scale}} of {{Dynamic Heterogeneities}} at the {{Glass Transition Determined}} by {{Multidimensional Nuclear Magnetic Resonance}}},
  author = {Tracht, U. and Wilhelm, M. and Heuer, A. and Feng, H. and {Schmidt-Rohr}, K. and Spiess, H. W.},
  year = {1998},
  month = sep,
  journal = {Physical Review Letters},
  volume = {81},
  number = {13},
  pages = {2727--2730},
  issn = {0031-9007, 1079-7114},
  doi = {10.1103/PhysRevLett.81.2727},
  urldate = {2025-10-27},
  copyright = {http://link.aps.org/licenses/aps-default-license},
  langid = {english}
}

@article{weeksThreeDimensionalDirectImaging2000,
  title = {Three-{{Dimensional Direct Imaging}} of {{Structural Relaxation Near}} the {{Colloidal Glass Transition}}},
  author = {Weeks, Eric R. and Crocker, J. C. and Levitt, Andrew C. and Schofield, Andrew and Weitz, D. A.},
  year = {2000},
  month = jan,
  journal = {Science},
  volume = {287},
  number = {5453},
  pages = {627--631},
  issn = {0036-8075, 1095-9203},
  doi = {10.1126/science.287.5453.627},
  urldate = {2025-10-27},
  langid = {english}
}

@article{whitakerKineticStabilityHeat2015,
  title = {Kinetic Stability and Heat Capacity of Vapor-Deposited Glasses of {\emph{o}} -Terphenyl},
  author = {Whitaker, Katherine R. and Tylinski, M. and Ahrenberg, Mathias and Schick, Christoph and Ediger, M. D.},
  year = {2015},
  month = aug,
  journal = {The Journal of Chemical Physics},
  volume = {143},
  number = {8},
  pages = {084511},
  issn = {0021-9606, 1089-7690},
  doi = {10.1063/1.4929511},
  urldate = {2025-10-27},
  langid = {english}
}

@article{xuLowfrequencyVibrationalDensity2024b,
  title = {Low-Frequency Vibrational Density of States of Ordinary and Ultra-Stable Glasses},
  author = {Xu, Ding and Zhang, Shiyun and Tong, Hua and Wang, Lijin and Xu, Ning},
  year = {2024},
  month = feb,
  journal = {Nature Communications},
  volume = {15},
  number = {1},
  pages = {1424},
  issn = {2041-1723},
  doi = {10.1038/s41467-024-45671-8},
  urldate = {2025-10-27},
  langid = {english}
}

@article{zhangEmergentFacilitationBehavior2017,
  title = {Emergent Facilitation Behavior in a Distinguishable-Particle Lattice Model of Glass},
  author = {Zhang, Ling-Han and Lam, Chi-Hang},
  year = {2017},
  month = may,
  journal = {Physical Review B},
  volume = {95},
  number = {18},
  pages = {184202},
  issn = {2469-9950, 2469-9969},
  doi = {10.1103/PhysRevB.95.184202},
  urldate = {2025-10-27},
  copyright = {http://link.aps.org/licenses/aps-default-license},
  langid = {english}
}

\section*{Data Availability}
The data that support the findings of this study are available from the corresponding author upon reasonable request.

\appendix
\section{Simulation Setups}\label{ap:simulation}
We perform molecular dynamics simulations
of the three-dimensional ($d=3$)
Kob--Andersen model~\cite{kobTestingModecouplingTheory1995e}
in this study.
This model is a canonical supercooled-liquid model
inspired by the metallic glass
$\mathrm{Ni}_{80}\mathrm{P}_{20}$,
and is described by a simple Lennard--Jones potential,
$V_{ij}=4\epsilon_{ij}\left[(\sigma_{ij}/r_{ij})^{12}-(\sigma_{ij}/r_{ij})^{6}\right]$,
where $r_{ij}$ denotes the distance
between particles $i$ and $j$,
$\epsilon_{ij}$ sets the interaction energy scale,
and $\sigma_{ij}$ determines the interaction range
between the two particles.
In the KAM,
these parameters are set in a non-additive manner as
$\sigma_{AA}=1$, $\sigma_{BB}=0.88$, $\sigma_{AB}=0.8$,
$\epsilon_{AA}=1$, $\epsilon_{BB}=0.5$, and $\epsilon_{AB}=1.5$.
Here, the subscripts $A$ and $B$
distinguish the two particle species.
The cutoff distance is set to
$r_{ij}^{\rm c}=2.5\sigma_{ij}$,
and the number density is fixed to
$\rho=N/V=1.2$,
where $V$ and $N$ denote the system volume
and the total number of particles, respectively.
The number ratio of the particle species is
$N_A:N_B=80:20$.
The particle mass is set to $m$
for both species.
In this Letter,
all physical quantities are nondimensionalized
using the energy unit $\epsilon_{AA}$,
the length unit $\sigma_{AA}$,
and the mass unit $m$.

In this study,
we analyze configurations located at saddle points
of the potential energy landscape.
To obtain such saddle-point configurations,
it is necessary to smooth the interaction potential
so that it continuously goes to zero
up to the second derivative
at the cutoff distance $r_{ij}^{\rm c}$.
Following Refs.~\cite{grigeraGeometricApproachDynamic2002b,
coslovichLocalizationTransitionUnderlies2019b},
we smooth the potential using a cubic polynomial of the form
\begin{align}
V_{ij}^{\rm cubic}=V_{ij}+B_{ij}(a_{ij}-r_{ij})^3+C_{ij}.
\label{eq:cubic}
\end{align}
Here, we introduce
\begin{align}
a_{ij}&=r_{ij}^{\rm c}-2\frac{V_{ij}^\prime(r_{ij}^{\rm c})}{V_{ij}^{\prime\prime}(r_{ij}^{\rm c})},\\
B_{ij}&=\frac{\left[V_{ij}^{\prime\prime}(r_{ij}^{\rm c})\right]^2}{12V_{ij}^\prime(r_{ij}^{\rm c})},\\
C_{ij}&=\frac{\left[V_{ij}^{\prime\prime}(r_{ij}^{\rm c})\right]^3}{216B_{ij}^2}-V_{ij}(r_{ij}^{\rm c}),
\end{align}
where $V_{ij}^\prime(r_{ij})\equiv\partial V_{ij}/\partial r_{ij}$ and
$V_{ij}^{\prime\prime}(r_{ij})\equiv\partial^2 V_{ij}/\partial r_{ij}^2$.

Since the number density is fixed in the KAM, only the system size $N$ and the temperature $T$ are the free parameters.
In this study, we consider the ranges $200\le N\le 1500$ and $0.44\le T\le 1.0$.
We note that this temperature range lies above the MCT point $T_{\rm MCT}\approx 0.435$.
For each system size, simulations were initialized from completely random particle configurations, and the calculations were started at the so-called onset temperature $T_{\rm onset}\approx 1.0$~\cite{sastrySignaturesDistinctDynamical1998b}, at which slow dynamics begin to emerge.
Subsequently, the temperature was decreased sequentially by performing simulations at lower temperatures, using the final configuration at each temperature as the initial configuration for the subsequent simulation.
At each temperature, after performing equilibration run with durations longer than $20$ times the $\alpha$-relaxation time, we carried out production runs of the same length.
To obtain reliable statistical averages, simulations were performed for at least 256 independent samples for each combination of parameters $N$ and $T$.
The time step of the molecular dynamics simulations is fixed at $\Delta t = 0.005$.
The relaxation-time parameter of the Nosé--Hoover thermostat is set to $\tau_T = 50\Delta t$, following Ref.~\cite{coslovichDynamicThermodynamicCrossover2018a}.

Several studies has discussed that, even in the KAM, the crystallization becomes non-negligible at very low temperatures~\cite{coslovichDynamicThermodynamicCrossover2018a,ingebrigtsenCrystallizationInstabilityGlassForming2019a,dasCrossoverDynamicsKobAndersen2022a,ortliebProbingExcitationsCooperatively2023}.
However, within the range of $T$ and $N$ considered in our study, only few crystallization samples are detected and they played almost no role for our observables.
Therefore, we did not exclude any samples from the analyses.
We summarize how the crystallization affect the results quantitatively in the accompanying full paper~\cite{DHFull}.

\section{Finite size scaling}\label{ap:fss}
Assuming the scaling relations introduced in the main text,
\begin{align}
    \xi &\sim T^{-\nu}, \label{eq:gamma} \\
    \chi_4^\ast &\sim T^{-\gamma},
\end{align}
we expect that a plot of $\chi_4^\ast L^{-\gamma/\nu}$ as a function of $T L^{1/\nu}$ for different system sizes $N$ exhibits a collapse of all data onto a single master curve.
Below, we explain the rationale behind this expectation.

To account for the effects of system-size-dependent crossover on the scaling hypothesis in Eq.~\eqref{eq:gamma}, we introduce a scaling function $f(x)$ and write
\begin{align}
    \chi_4^\ast(N,T)=T^{-\gamma}f\!\left(T / T_{\rm FS}(N)\right).
    \label{eq:scaling_func}
\end{align}
Here, $T_{\rm FS}(N)$ denotes a system-size-dependent crossover temperature at which the results for each system size start to deviate from the power-law behavior.
We explicitly indicate the dependence of
$\chi_4^\ast$ and $T_{\rm FS}$
on $N$ and $T$ through their arguments.
Such finite-size effects are expected to emerge
when the system size becomes comparable to
the correlation length,
which leads to the scaling relation
$T_{\rm FS} \sim L^{-1/\nu}$.
Substituting this relation into Eq.~\eqref{eq:scaling_func} and performing a change of variables so that the right-hand side is expressed solely in terms of $T/T_{\rm FS}=T L^{1/\nu}$, we obtain
\begin{align}
    \chi_4^\ast=L^{\gamma/\nu}(T L^{1/\nu})^{-\gamma}f(T L^{1/\nu}).
\end{align}
From this result, we expect that a plot of $\chi_4^\ast L^{-\gamma/\nu}$ as a function of $T L^{1/\nu}$ will lead to a collapse of the data for all system sizes, including the crossover regime.

\section{Time scale measurement}\label{ap:taux}
We measure the $\alpha$ relaxation time $\tau_\alpha$ using the following two-mode relaxation model:
\begin{align}
    Q_{\rm 2m}(t)\equiv(1-f_{\rm c})\exp\left[-(t/t_{\rm s})^2\right]+f_{\rm c}\exp\left[-(t/\tau_\alpha)^{\beta_{\rm KWW}}\right].\label{eq:2mode}
\end{align}
The first term on the right-hand side of Eq.~\ref{eq:2mode} represents the contribution from the fast mode.
Following Ref.~\cite{dasCrossoverDynamicsKobAndersen2022a}, we assume a Gaussian form for this mode and fix the exponent to two.
For the slow mode, which appears as the second term, we assume a Kohlrausch--Williams--Watts (KWW)-type stretched exponential form.
In this two-mode relaxation model, there are four parameters: the fast-mode characteristic time $t_{\rm s}$, the slow-mode weight $f_{\rm c}$, the slow-mode relaxation time $\tau_\alpha$, and the stretching exponent $\beta_{\rm KWW}$, which characterizes the shape of the slow-mode relaxation function.
We estimate these four parameters, including $\tau_\alpha$, as functions of temperature and system size by fitting the measured $Q(t)$ to Eq.~\eqref{eq:2mode}.
In the main text, we use the values of $\tau_\alpha$ determined by this procedure.
The detailed dependence of all four parameters on temperature and system size is reported in the accompanying full paper~\cite{DHFull}.

Regarding the determination of $\tilde{\tau}_4$, we interpolate the measured data points of $\tilde{\chi}_4$, since the logarithmic time sampling employed here does not allow a precise identification of the peak position.
In this work, we use cubic spline interpolation and identify the peak position of the resulting interpolated curve as $\tilde{\tau}_4$.
The values of $\chi_4^\ast$ and $\tilde{\chi}_4^\ast$ reported in the main text are likewise determined as the peak heights of the same interpolated curve.

\end{document}


\title{
{Zero-temperature Avalanche Criticality Governing \\Dynamical Heterogeneity in Supercooled Liquids} \\--- supplemental material
}

\maketitle

\section{Binder parameter measurement}\label{sec:binder}
\begin{figure*}[hb]
\includegraphics[width=0.5\linewidth]{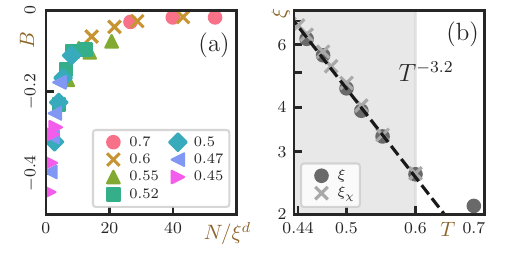}
\caption{
(a) Linear plot of the Binder parameter computed from our simulation results as a function of the scaled system size $N/\xi^d$, where $\xi$ is taken from Ref.~\cite{karmakarGrowingLengthTime2009a}.
(b) Log–log plots of the two correlation lengths, $\xi$ and $\xi_\chi$, as functions of the temperature $T$.
The dashed line indicates a power-law fit to $\xi$ in the scaling regime.
The shaded region indicates the scaling regime $T \le T_{\rm ava}$.
The same information as that shown in Fig.~3(a) in the main text is presented.
\label{fig:binder}}
\end{figure*}

In this section, we present the temperature ($T$) and system-size ($N$) dependence of the Binder parameter
\begin{align}
    B\equiv \frac{\langle[q(\tau_4)-\langle q(\tau_4)\rangle]^4\rangle}{3\langle [q(\tau_4)-\langle q(\tau_4)\rangle]^2\rangle^2}-1
\end{align}
obtained from our simulation results.
In particular, Fig.~\ref{fig:binder}(a) shows $B$ plotted as a function of the scaled system size $N/\xi^d$, where $\xi(T)$ is taken from the correlation length reported in Ref.~\cite{karmakarGrowingLengthTime2009a}.
Consistent with Ref.~\cite{karmakarGrowingLengthTime2009a}, the data for different temperatures and system sizes collapse onto a single master curve.
For reference, Fig.~\ref{fig:binder}(b) shows the values of the correlation length $\xi$ at each temperature (Fig.~\ref{fig:binder}(b) is identical to Fig.~3(a) in the main text).

\newpage
\section{Finite size scaling based mode-coupling transition ansatz}\label{sec:MCT}
\begin{figure*}[hb]
\includegraphics[width=0.8\linewidth]{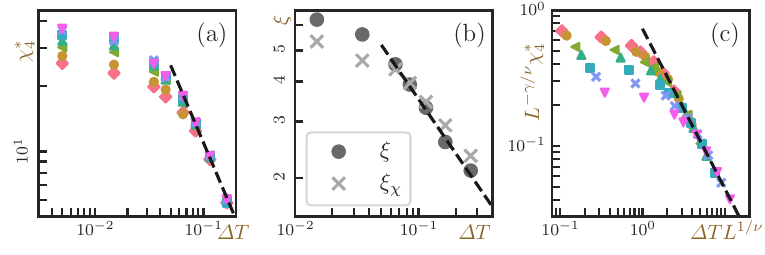}
\caption{
(a) Log–log plot of $\chi_4^\ast$ as a function of the distance from the MCT point, $|T - T_{\rm MCT}|$.
Different symbols correspond to different system sizes, as in Fig.~2 of the main text.
The dashed line indicates the power-law fit to $\chi_4^\ast$ in the temperature range where apparent power-law behavior is observed.
(b) Log–log plots of the two correlation lengths, $\xi$ and $\xi_\chi$, as functions of the temperature $T$.
The dashed line indicates the power-law fit to $\xi$ in the temperature range where apparent power-law behavior is observed.
(c) Finite-size scaling of $\chi_4^\ast$ assuming a criticality ansatz with the MCT point as the critical point.
The dashed line represents the same power-law as in panel (a).  \label{fig:MCT}}
\end{figure*}

For the Kob–Andersen model, the mode-coupling transition temperature, $T_{\rm MCT}$, has been estimated from numerical results to be $T_{\rm MCT} \approx 0.435$ \cite{kobTestingModecouplingTheory1995e}.
Mode-coupling theory predicts a power-law divergence of slow dynamics at this critical point.
Assuming criticality with the MCT point as the critical point, we plot both the dynamical susceptibility and the dynamical correlation length as functions of the distance from the critical point, $\Delta T \equiv |T - T_{\rm MCT}|$, as shown in Fig.~\ref{fig:MCT}(a,b).
Indeed, both observables exhibit apparent power-law behavior over a certain temperature range.
However, compared to the zero-temperature avalanche criticality picture presented in the main text, the power-law regime is confined to a narrower range at higher temperatures. 
Even for our largest system size, $N=1500$, clear deviations from the power-law behavior are observed already at temperatures as high as $T \approx 0.5$, well above $T_{\rm MCT}$.
Moreover, no clear criterion for identifying the temperature range governed by criticality has been established, at least at present. Therefore, in the following, we determine the critical exponents by direct fitting within the scaling regime identified by visual inspection.

By fitting $\chi_4^\ast$ and $\xi$ to power laws within the temperature range where they exhibit apparent power-law behavior, and performing finite-size scaling using the resulting exponents ($\nu = 0.56$, $\gamma = 1.18$), we obtain the result shown in Fig.~\ref{fig:MCT}(c), where data for different system sizes do not collapse onto a single curve.
Furthermore, the correlation length $\xi_\chi$ estimated from the relation $\xi_\chi = \chi_4^{\ast,1/d_f}$ using the obtained exponents does not agree with $\xi$ [Fig.~\ref{fig:MCT}(b)].
We also verified that varying $T_{\rm MCT}$ within the range $0.43 \le T_{\rm MCT} \le 0.44$ does not qualitatively affect the behaviors observed in Figs.~\ref{fig:MCT}(b) and \ref{fig:MCT}(c).

These results indicate that a criticality scenario based on the MCT point fails to account for the observed behavior, at least when the well-established dynamical correlation length is used as the critical correlation length.

\newpage
\section{Unscaled displacement-based dynamical susceptibility peak $\tilde{\chi}_4^\ast$}\label{sec:chi4tilde}
In Fig. 5(a) of the main text, we presented only the results after finite-size scaling for the temperature dependence of the displacement-based dynamical susceptibility, $\tilde{\chi}_4^\ast$. Here, we also show the unscaled results for comparison, as in Fig. 2.

\begin{figure*}[hb]
\includegraphics[width=0.5\linewidth]{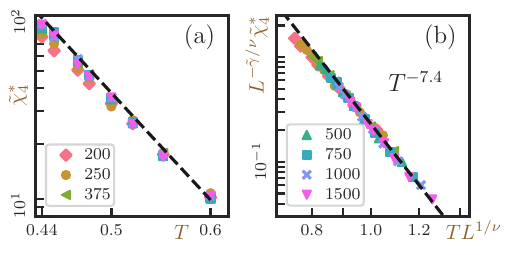}
\caption{
Log--log plots of the temperature and system-size dependence of the
displacement-based dynamical susceptibility peak $\tilde{\chi}_4^\ast$ in the scaling regime
$T \le T_{\rm ava} \approx 0.6$.
(a) Raw data.
(b) Finite-size scaling results.
Symbols represent different system sizes, as indicated in the legend.
The dashed line shows the power-law scaling
$\tilde{\chi}_4^\ast \sim T^{-\tilde{\gamma}}$.
  \label{fig:tilde_chi4}}
\end{figure*}

\bibliography{DH_INMs}